\documentclass[fleqn,twoside]{article}
\usepackage{espcrc2}

\usepackage{psfrag}

\usepackage{graphicx}
\usepackage[figuresright]{rotating}

\newcommand{\AmS}{{\protect\the\textfont2
  A\kern-.1667em\lower.5ex\hbox{M}\kern-.125emS}}

\def\pel{{(\ell)}}
\def\pel{{(\ell)}}
\def\cG{  {\cal G}  }
\def\cO{  {\cal O}  }

\title{
Scattering\hspace{0.14cm}amplitudes\hspace{0.14cm}on\hspace{0.14cm}the\hspace{0.14cm}Coulomb\hspace{0.14cm}branch\hspace{0.14cm}of\hspace{0.14cm}$\mathcal{N}=4$\hspace{0.14cm}super\hspace{0.14cm}Yang-Mills\thanks{HU-EP-10/22. Talk given at {\it Integrability in Scattering Amplitudes}, IAS, April 2010, 
{\it Loops and Legs in Quantum Field Theory}, DESY, April 2010 and
{\it Amplitudes 2010}, QMUL, May 2010.}
}

\author{J. M. Henn\address[MCSD]{Humboldt Universit\"{a}t zu Berlin, 
       Newtonstr. 15, \\ 
        12489 Berlin, Germany}%
                }
       
\begin{document}

\begin{abstract}
We discuss planar scattering amplitudes on the Coulomb branch of $\mathcal{N}=4$ super Yang-Mills.
The vacuum expectation values on the Coulomb branch can be used to regulate infrared divergences.
We argue that this has a number of conceptual as well as practical advantages over dimensional regularisation. 
\vspace{1pc}
\end{abstract}

\maketitle

\section{INTRODUCTION}

Scattering amplitudes in $\mathcal{N}=4$ super Yang-Mills (SYM) have many 
surprising properties, especially in the planar limit. 
Being a massless gauge theory, soft and collinear 
infrared divergences appear that have to be regulated.
After a suitable removal
of the universal divergent part
one can define a finite part. 
The latter was found to be surprisingly simple, initially
perturbatively \cite{Anastasiou:2003kj,Bern:2005iz} and
later, via the conjectured AdS/CFT correspondence,  
at strong coupling  \cite{Alday:2007hr}.
Its form (and that of the five-point amplitude) is  
believed to be known to all orders in 
the coupling constant \cite{Drummond:2007au}
(for reviews see \cite{Alday:2008yw,Henn:2009bd}).

Given these impressive developments, one may hope
that all-order results can also be derived for 
amplitudes with six and more external particles, at least
in the planar limit.
Having this motivation in mind, it makes sense to try
to use similar methods at weak and strong coupling.
An apparent dissimilarity so far is that at weak coupling,
the use of dimensional regularisation is predominant,
while at strong coupling a different regulator seems to be
more natural, where one considers open strings attached 
to $D$-branes at a non-zero separation 
\cite{Alday:2007hr,Kawai:2007eg,McGreevy:2008zy,Berkovits:2008ic}.

In the dual field theory, this corresponds to going to the 
Coulomb branch, and was considered in 
\cite{Schabinger:2008ah,Alday:2009zm}.
We take $\mathcal{N}=4$ SYM with gauge group $U(N+M)$ and 
break it to $U(N) \times U(M)$ by means of a Higgs mechanism. 
This gives rise to massive particles in the broken part of the 
gauge group, while the particles in
the unbroken $U(N)$ and $U(M)$ parts remain massless.
If we scatter particles with labels in the $U(M)$ part of the gauge group we can select
diagrams where the particles in the loop(s) are in the $U(N)$ part by taking $N \gg M$.
In this way, we arrive at the following situation: 
The scattered particles are massless, and couple via particles of mass $m$ that travel 
along the perimeter of a given diagram, making the integrals infrared finite. 

We will now discuss various properties of the
massive amplitudes by means of a one-loop example.
We then show how the mass regulator can be used to
rederive results obtained previously in dimensional 
regularisation \cite{Bern:2005iz,Bern:2006ew,Cachazo:2006az}
in a simpler way \cite{Henn:2010bk,Henn:2010ir}. 
We also show that the mass-regulated integrals behave 
nicely in the Regge limit, and identify
a class of integrals giving the correct leading 
log and next-to leading log
contributions to all orders in the coupling constant.

\section{ONE-LOOP EXAMPLE}

Let us illustrate some features of the massive amplitudes by means of a one-loop example.
We denote the (colour-ordered) four-point amplitude, normalised by the tree-level contribution, by $M_{4}$.
Its expansion in the `t Hooft coupling $a = g^2 N/(8 \pi^2 )$ reads
$M_{4} = 1 + a \, M_{4}^{(1)} + \cO(a^2)$.
The one-loop contribution $M_{4}^{(1)}$ is given by  a scalar box integral \cite{Alday:2009zm},
with massless external legs and a uniform mass $m$ in the loop.

It is noteworthy that its exact expression in $m$ involves logarithms and dilogarithms only.
Let us compare this to dimensional regularisation, where the result is a hypergeometric 
function which depends on $D=4-2\epsilon$, with $\epsilon<0$. When expanding the latter in $\epsilon$, 
one obtains higher transcendental functions, see e.g. equation (B.2) in \cite{Bern:2005iz}.

Coming back to the Coulomb branch, in practice we will not need $M_{4}$ in its full generality. 
There are various interesting limits that one can consider, as we discuss presently.

\subsection{Regge limit}
We use the usual notation $s=(p_{1}+p_{2})^2$ and $t=(p_{2}+p_{3})^2$ for the Mandelstam
invariants. In the Regge limit $s \gg t,m^2 $, we obtain
\begin{equation}
M_{4}^{(1)} = \log \frac{s}{m^2} \, \alpha\left(\frac{t}{m^2}\right) + \cO(s^0 ) \,,
\end{equation}
where $(\alpha+1)$ is the Regge trajectory.
As we will see, this limit organises integrals contributing to the amplitude in a systematic way. 
We will also give a simple way of seeing that a single
logarithm in $s$ appears per loop order. 

\subsection{Low energy limit}  
The limit $m^2 \gg s,t$ was discussed in \cite{Gorsky:2009dr}.

\subsection{Small mass limit} 
Here we take $m^2 \ll s,t$, thereby approaching the massless theory,
with the infrared (IR) divergences regulated by $m^2$.
In our one-loop example, we obtain 
\begin{eqnarray}\label{M4smallmass}
\nonumber  M_{4}^{(1)}  
            &=& - \left[  \frac{1}{2} \log^2  \frac{m^2 }{s}  +  \frac{1}{2} \log^2  \frac{m^2}{t}  \right] \\
            &&  +\frac{1}{2} \log^2  \frac{s}{t}  + \frac{\pi^2}{2} + \cO(m^2)\,,
\end{eqnarray}
where we have written the result in a way as to make the origin of the $\log^{2} m^2$ terms manifest. 
As $m^2 \to 0$, we obtain double logarithms per loop order as a result of soft and collinear divergences.
These logarithms are the analogs of $(\mu^2 / s)^{\epsilon}/\epsilon^2$ in dimensional regularisation, where one has 
\begin{eqnarray}\label{M4dimreg}
\nonumber  M_{4}^{(1)}   & =&  - \left[  \frac{1}{\epsilon^2 }  \left( \frac{\mu^2}{s} \right)^{\epsilon} +  \frac{1}{\epsilon^2 }\left( \frac{\mu^2}{t} \right)^{\epsilon} \right] \\
&& +\frac{1}{2} \log^2  \frac{s}{t}  + \frac{2\pi^2}{3} + \cO(\epsilon)\,.
\end{eqnarray}
The IR divergences appearing in (\ref{M4smallmass}) and (\ref{M4dimreg}), 
which take the form of $\log^2 m^2 $ and $1/\epsilon^2$, respectively, are
well understood. We will be interested in the finite part of (the logarithm of) 
$M_{4}$, which is scheme-independent up to an additive constant.

\subsection{Geometrical interpretation}
Interestingly, the one-loop box integrals can be interpreted as polytopes in AdS${}_{5}$ 
\cite{Mason:2010pg} (see also \cite{Davydychev:1997wa}).
Moreover, the mass regulator is natural in the context of
momentum twistor space, see \cite{Hodges:2010kq}. This might be important
when trying to extend the approach of \cite{ArkaniHamed:2009dn} to loop level integrals.

\section{HIGHER LOOP ORDERS AND\\ EXPONENTIATION}

Let us now review the structure of higher-loop corrections to 
scattering amplitudes. In a generic gauge theory 
one can write the planar amplitude in the following factorised form 
(see \cite{Bern:2005iz} and references therein)
\begin{eqnarray}\label{IRstructure}
\log M_{4} = D(s) + D(t) + F_{4}\left(\frac{s}{t} \right) + \cO(\epsilon, m^2) \,,
\end{eqnarray}
where the r.h.s. is a sum of IR divergent terms $D$ and a finite term $F_{4}$.
If the $\beta$ function is zero, as in our case, the explicit form of $D(s)$ is particularly simple. 
In dimensional regularisation,
\begin{eqnarray}
\label{altbds}
D(s) =
-\frac{1}{2} \, \sum_{\ell \ge 1}  a^\ell
\left[ 
\frac{\Gamma_{\rm cusp}^\pel}{  (\ell \epsilon)^2} +
\frac{\cG_0^\pel} {( \ell \epsilon )} 
\right]
\left( \frac{\mu^2 }{ s}\right)^{\ell \epsilon} \,. 
\end{eqnarray}
Here $\Gamma_{\rm cusp}^\pel$ are the expansion coefficients of the universal
cusp anomalous dimension \cite{Korchemskaya:1992je}, and  $\cG_0^\pel $
those of the scheme-dependent collinear anomalous dimension.

The structure of the infrared singular terms when using the mass regulator is \cite{Korchemsky:1988hd,Mitov:2006xs}
\begin{eqnarray} \label{IRmassdivs}
D(s) =   -\frac{1}{4} \Gamma_{ {\rm cusp}}(a)    \log^2  \frac{s}{m^2}  -  \tilde{\cG}_{0}(a)   \log  \frac{s}{m^2}  \,,
\end{eqnarray}
where $ \tilde{\cG}_{0}$ is the analog of ${\cG}_{0}$.
The infrared terms $D$ being universal, we are interested in the
finite part $F_{4}$ (which is equal in both schemes up to a 
coupling-dependent constant).
Surprisingly, the latter also turns out to be very 
simple \cite{Anastasiou:2003kj,Bern:2005iz,Alday:2007hr,Drummond:2007au},
\begin{eqnarray}\label{BDS4point}
F_{4}=  \frac{1}{4}{\Gamma_{\rm cusp}(a)}   \, \log^2 {  \frac{s}{t} } 
+ {\rm const}(a) \,.
\end{eqnarray}
Note that when computing $F_{4}$ perturbatively in dimensional regularisation, 
e.g. to two-loop order,
equation (\ref{IRstructure}) implies that one needs to know 
\begin{equation}\label{two-loop-exp}
M_{4}^{(2)} - \frac{1}{2} \left( M^{(1)}_{4} \right)^2
\end{equation} 
up to terms of $\cO(\epsilon)$. 
However, $1/\epsilon$ terms in $M^{(1)}_{4}$ lead to an 
interference of the type $1/\epsilon \times \cO(\epsilon) = \cO(1)$, 
and therefore the $\cO(\epsilon)$ and $\cO(\epsilon^2)$ terms in $M^{(1)}_{4}$
need to be computed too. Likewise, at each new loop order, 
all lower-loop results need to be 
extended by two additional orders in $\epsilon$.

Let us now consider equation (\ref{two-loop-exp}) in mass 
regularisation \cite{Alday:2009zm,Henn:2010bk}.
There, we can drop terms that vanish
as the regulator goes to zero. This is because 
we have $\cO(m^2 ) \times \log m^2 \to 0 $.
This trivial observation has very important practical consequences.
In particular, for a calculation at arbitrary loop order, the only information
needed about $M^{(1)}_{4}$ is already given in equation (\ref{M4smallmass}).

\section{EXTENDED DUAL CONFORMAL SYMMETRY}

There is a lot of evidence by now that planar scattering amplitudes in $\mathcal{N}=4$ SYM 
possess a dual (super)conformal symmetry \cite{Drummond:2006rz,Drummond:2008vq}.
The symmetry acts in a dual space
\begin{eqnarray}
p^{\mu}_{i} = x^{\mu}_{i} - x^{\mu}_{i+1}\,,
\end{eqnarray}
where the cyclicity condition $x_{i+n}^{\mu} \equiv x_{i}^{\mu}$, with $n$ 
being the number of scattered particles, is tacitly implied.
It can be seen that the loop integrand of e.g. the one-loop box integral
has a symmetry under inversions in the dual space, or equivalently,
under special conformal transformations,
\begin{eqnarray}
K^{\mu} =  \sum_{i=1}^{n} \left[  2 x_{i}^{\mu} \, {x^{\nu}_i \frac{\partial}{\partial {x_{i}^{\nu}}}} - {x_{i}^2} \frac{\partial}{\partial x_{i \mu}} \right]\,.
\end{eqnarray}
This symmetry is broken in dimensional regularisation,
which is why integrals whose integrand naively has the symmetry were dubbed ``pseudo-conformal''.
The breaking is believed to be under control on the level of the amplitudes, which
are conjectured to satisfy anomalous Ward identities \cite{Drummond:2008vq} initially derived for certain Wilson loops \cite{Drummond:2007au}.

We will now argue that the symmetry can be repaired at the level of the loop integrals
by refining the Higgs setup discussed above \cite{Alday:2009zm}.
Let us further break the gauge symmetry from 
$U(N) \times U(M)$
to $U(N) \times U(1)^{M}$. 
In this way, we introduce several particle masses.
This changes the on-shell conditions of the scattered 
particles from $p_{i}^2 = 0$ to $p_{i}^2 =  - (m_{i} - m_{i+1})^2$
and gives different masses to propagators in the loop,
\begin{eqnarray}
I =  \hat{x}_{13}^2 \, \hat{x}_{24}^2  \, \int  \, \frac{d^{4}x_{a}  }{ \prod_{i=1}^{4} (x_{ia}^2 + m_{i}^2) }  \,,
\end{eqnarray}
where $\hat{x}_{ij}^2 = x_{ij}^2 + (m_{i}-m_{j})^2 $.
The integral is finite, and moreover it has a symmetry, 
provided that we act on the 
masses as well as the dual coordinates,
\begin{eqnarray}\label{extendedconformal}
 \hat{K}^{\mu} I = 0 \,,
\end{eqnarray}
where
\begin{eqnarray}\label{extendedconformalgenerator}
  \hat{K}^{\mu} = K^{\mu} + \sum_{i=1}^{n} \left[  2 x_{i}^{\mu} \, {m_i \frac{\partial}{\partial {m_{i}}}} - {m_{i}^2} \frac{\partial}{\partial x_{i \, \mu}} \right]\,.
\end{eqnarray}
We call $\hat{K}^{\mu}$ {\it extended} dual conformal transformations.

This has a natural interpretation in string theory. There, the
dual conformal symmetry is viewed as the isometry group of a (T-dual) AdS${}_{5}$ space \cite{Alday:2007hr,Berkovits:2008ic},
with $(x^{\mu}, m)$ being its Poincar\'{e} coordinates.
Usually, when discussing symmetries in the boundary field theory, one sets
the radial coordinate $m=0$. Here, we simply use a different realisation 
of the SO(2,4) symmetry with $m \neq 0$, which leads to the generator (\ref{extendedconformalgenerator}) 
 (see also \cite{Jevicki:1998qs} for a similar discussion of conformal symmetry).

A comment is that from a conventional viewpoint one might be reluctant to
call (\ref{extendedconformal}) a symmetry since the transformations relate scattering 
amplitudes with particles having different masses. However, from the 
AdS perspective the $m_{i}$ are regarded as coordinates, just as the dual coordinates $x_{i}$.
Whichever interpretation one prefers, the 
upshot is that the amplitudes satisfy the conjectured differential equation $\hat{K}^{\mu} M = 0$.
As we will see presently, this has important implications for the loop-level integral basis.
Also, those integrals can depend on the kinematical variables only in a specific way.
For example, in the four-point case, they can be functions of the following two conformally invariant
variables only  \cite{Alday:2009zm},
\begin{eqnarray}
u &=& \frac{m_{1} m_{3}}{x_{13}^2 + (m_{1}-m_{3})^2}\,, \label{defu}\\
v &=& \frac{m_{2} m_{4}}{x_{24}^2 + (m_{2}-m_{4})^2}\,,\label{defv}
\end{eqnarray}
where we recall that $x_{13}^2 = s$ and $x_{24}^2 = t$.

\section{HIGHER LOOP INTEGRAL BASIS}
It has been observed and discussed in many papers 
\cite{Drummond:2006rz,Drummond:2007aua,Nguyen:2007ya,Bern:2007ct,Bern:2008ap} 
that the loop integrals contributing
to amplitudes in $\mathcal{N}=4$ SYM are of a very restricted set. The concept
of ``pseudo-conformal'' integrals has been a useful one, although a precise
definition of which integrals are pseudo-conformal and which ones are not is
difficult due to the issue of IR divergences. One improvement in the
setup we propose is that there is a clear definition of integrals invariant under
extended dual conformal transformations, namely equation (\ref{extendedconformal}).

It is natural to speculate that at a given loop order $L$, the amplitude can be 
written as a linear combination of integrals $I$ satisfying (\ref{extendedconformal}),
with certain coefficients $c(I)$,
\begin{eqnarray}\label{dualconfansatz}
M^{(L)} = \sum_{I} \, c(I) \, I \,.
\end{eqnarray}
It is obviously very useful to know in advance which integrals can appear in a
calculation, as is the case at one-loop order.

In practice it is easy to write down all dual conformal integrals at a given
loop order and number of external particles (e.g. using dual graphs). 
We remark that an implication of the symmetry is that all triangle subgraphs
be absent. At one loop, this necessary requirement was shown to hold 
\cite{Boels:2010mj} (see also \cite{Schabinger:2008ah}).

If the dual conformal ansatz is to reproduce the scattering amplitude $M^{(L)}$, it must
have the correct infrared structure, which in the equal mass case is dictated by equations
(\ref{IRstructure}) and (\ref{IRmassdivs}). In general this implies relations between the 
coefficients $c(I)$ appearing in (\ref{dualconfansatz}). As we will see, these
IR consistency equations are in some cases sufficient to determine the $c(I)$.

The two-loop four-point amplitude serves as an illustration of the last statement, 
since only a single integral is allowed by dual conformal symmetry, whose normalisation
is fixed by the exponentiation of IR divergences \cite{Alday:2009zm}.
\begin{figure}
\begin{center}
\psfrag{p1}[cc][cc]{$p_{1}$}
\psfrag{p2}[cc][cc]{$p_{2}$}
\psfrag{p3}[cc][cc]{$p_{3}$}
\psfrag{p4}[cc][cc]{$p_{4}$}
\psfrag{I3a}[cc][cc]{$I_{3a}$}
\psfrag{I3b}[cc][cc]{$I_{3b}$}
\psfrag{I3c}[cc][cc]{$I_{3c}$}
\psfrag{I3d}[cc][cc]{$I_{3d}$}
\includegraphics[scale=0.4]{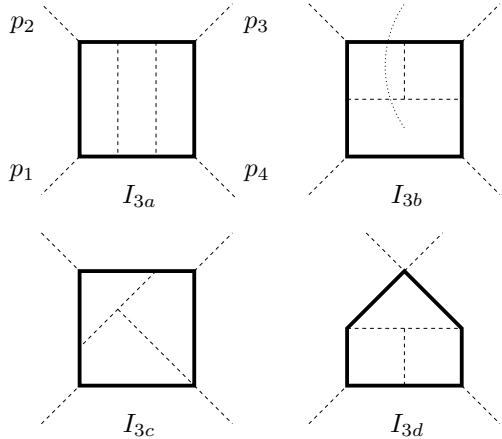}
\end{center}
\caption{Three-loop dual conformal integrals. Thick lines denote massive propagators,
dashed lines massless ones. $I_{3b}$ contains a loop-dependent numerator, as indicated by the dotted line.
The loop-independent normalisations $s^3 t\,, s t^2\,, m^2 s t$, and $m^2 t^2$ for $I_{3a}\,,I_{3b}\,,I_{3c}$ and $I_{3d}$, respectively, are not
shown.}
\label{fig:3loop}
\end{figure}
At three loops and four points, there are four dual conformal integrals as shown in Fig. \ref{fig:3loop}.
Therefore our ansatz (\ref{dualconfansatz}) becomes \cite{Henn:2010bk}
\begin{eqnarray}\label{Mthreeloop}
M_{4}^{(3)} &=& -\frac{1}{8} \, \left( c_{3a} I_{3a} +  c_{3b} I_{3b} + c_{3c} I_{3c} + c_{3d} I_{3d} \right) \nonumber \\
&& + \; \{ s \leftrightarrow t \} \,.
\end{eqnarray}
Let us for the moment set $s=t \,, \, L \equiv  \log({m^2}/{s})$ for simplicity. 
Then the infrared consistency equation, obtained by expanding (\ref{IRstructure}) and (\ref{IRmassdivs}) to
three-loop order and plugging in the known two-loop data, reads 
\begin{eqnarray}\label{MthreeloopIR}
M_{4}^{(3)} &=& -\frac{1}{6} L^6 + \frac{\pi^2}{12}  L^4 + 2 \zeta_{3} L^3 \nonumber\\ 
    && + \left(- \frac{\pi^4}{30}-\frac{\Gamma_{\rm cusp}^{(3)}}{4} \right) L^2 + \cO(L)\,.
\end{eqnarray}
On the other hand, we can explicitly compute the asymptotic expansion of the three-loop integrals.
To this end, we write down Mellin-Barnes representations of the integrals, 
and use the Mathematica packages {\it MB} \cite{Czakon:2005rk} and {\it MBasymptotics} \cite{MBasymptotics}
to perform the small $m^2$ expansion. We obtain
\begin{eqnarray}
I_{3a} &=& \frac{17}{90} L^6 + \frac{\pi^2}{9} L^4 + \cO(L^3) \,,  \\ 
 I_{3b} &=& \frac{43}{180} L^6 - \frac{2 \pi^2}{9} L^4 + \cO(L^3)  \,, \\ 
 I_{3c} &=& \cO(L^0) \,, \qquad I_{3d} = \cO(L) \,. 
 % I_{3c} &=& \cO(L^0) \,, \\
 %I_{3d} &=& \cO(L) \,. 
\end{eqnarray}
This obviously determines $c_{3a} =1$ and $c_{3b}=2$, while the 
coefficients $c_{3c}$ and $c_{3d}$ remain arbitrary.
We remark that in dimensional regularisation, one obtains the coefficients
$c_{3a} =1$, $c_{3b} =2$, $c_{3c} =0$ and $c_{3d} =0$ \cite{Bern:2005iz}.
It would be interesting to ascertain the values of the $c_{i}$ in the mass 
regulated setup by a direct computation.

We can nevertheless check the consistency of the ansatz (\ref{Mthreeloop}) with (\ref{BDS4point}).
Indeed, if $c_{3c}$ and/or $c_{3d}$ are non-zero, they could be accommodated by a 
change of the three-loop values of $\tilde{\cG}_{0}$ and 
the three-loop constant in the finite part of $F_{4}$.
We have checked numerically \cite{Henn:2010bk} for various values of $s \neq t$ that (\ref{Mthreeloop})
is in agreement with (\ref{BDS4point}).

We remark that while the coefficients of the logarithms in (\ref{MthreeloopIR}) are rather simple, 
the ones in the results for $I_{3a}$ and $I_{3b}$ are more complicated.
Perhaps a better way of performing the calculation exists which avoids the relative complexity
of the intermediate results.

In a recent paper, we have extended our computations of the four-point amplitude to the four-loop level \cite{Henn:2010ir}.
We numerically reproduce the known result for the four-loop value of the cusp anomalous dimension.
Perhaps it is a good illustration of the power of the mass regulator that we improve the numerical
accuracy of the initial computation \cite{Bern:2006ew} by five digits and that of the subsequent 
computation \cite{Cachazo:2006az} by two digits.

\section{REGGE LIMIT}

In the Regge limit $s \gg t, m^2$ one expects the four-point amplitude to have
the following form,
\begin{equation}\label{M4Regge}
M_{4}  = \beta(t/m^2 ) \left(\frac{s}{m^2}\right)^{\alpha(t/ m^2)} + \cO(s^0 )\,.
\end{equation}
Expanding (\ref{M4Regge}) in the coupling constant, at $L$
loops the leading and next-to leading logarithm (LL and NLL) will be $\log^L (s/m^2)$ and
$\log^{L-1} (s/m^2)$, respectively.

\begin{figure}
\begin{center}
\psfrag{p1}[cc][cc]{$p_{1}$}
\psfrag{p2}[cc][cc]{$p_{2}$}
\psfrag{p3}[cc][cc]{$p_{3}$}
\psfrag{p4}[cc][cc]{$p_{4}$}
\psfrag{a}[cc][cc]{(a)}
\psfrag{b}[cc][cc]{(b)}
\psfrag{dots}[cc][cc]{\dots}
\includegraphics[scale=0.55]{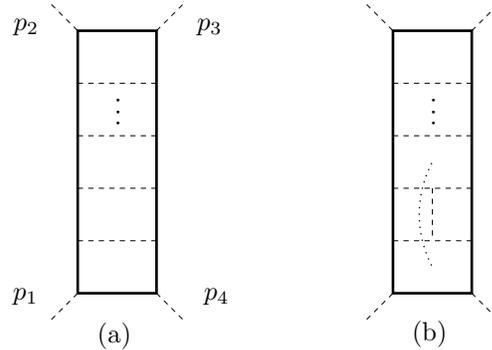}
\end{center}
\caption{Integrals dominating in the Regge limit of the four-point amplitude.  
The $L$-loop ladder (a) gives the correct LL term,
while at NLL only the subleading terms of (a) and the leading terms of (b)
are required, to all orders in the coupling constant.}
\label{fig:regge}
\end{figure}

We showed in \cite{Henn:2010bk,Henn:2010ir} that the LL and NLL terms 
can be obtained from a simple class of ladder diagrams and 
ladders with H-shaped insertions, see Fig. \ref{fig:regge}.
Note that this simple pattern is not present in dimensional
regularisation.

Dual conformal symmetry allows us to make an interesting observation \cite{Henn:2010bk}.
Recall that the four-point amplitude is a function of $u$ and $v$ defined in (\ref{defu}) and (\ref{defv}),
and the Regge limit implies $u \ll v$. 
If we set $m_{1} = m_{3} = m$ and
$m_{2} = m_{4} = M$, we have
\begin{equation}
u ={m^2}/{s} \,,\qquad v ={M^2}/{t} \,.
\end{equation}
and taking $m^2 \ll M^2$,  
we recover the Regge limit $u \ll v$ of the original scattering
process. However, our new setup corresponds to a ``Bhabha-like'' scattering process,
where heavy particles of approximate mass $M$ 
exchange lighter particles of mass $m$, at fixed $s$ and $t$.  
The absence of collinear divergences in such a process 
implies that only single logarithms in $u$ appear per loop order,
in agreement with (\ref{M4Regge}).

\section*{ACKNOWLEDGMENTS}
We are grateful to the organisers of the conferences where this talk
was given for their invitation and to the participants 
for many interesting discussions.
It is a pleasure to thank F.~Alday,  S.~Naculich, J.~Plefka, H.~Schnitzer, T.~Schuster and M.~Spradlin 
for collaboration on the topics presented here.


\begin{thebibliography}{20}

%\cite{Anastasiou:2003kj}
\bibitem{Anastasiou:2003kj}
  C.~Anastasiou, Z.~Bern, L.~J.~Dixon and D.~A.~Kosower,
  %``Planar amplitudes in maximally supersymmetric Yang-Mills theory,''
  Phys.\ Rev.\ Lett.\  {\bf 91} (2003) 251602
  [arXiv:hep-th/0309040].
  %%CITATION = PRLTA,91,251602;%%

%\cite{Bern:2005iz}
\bibitem{Bern:2005iz}
  Z.~Bern, L.~J.~Dixon and V.~A.~Smirnov,
  %``Iteration of planar amplitudes in maximally supersymmetric Yang-Mills
  %theory at three loops and beyond,''
  Phys.\ Rev.\  D {\bf 72} (2005) 085001
  [arXiv:hep-th/0505205].
  %%CITATION = PHRVA,D72,085001;%%

%\cite{Alday:2007hr}
\bibitem{Alday:2007hr}
  L.~F.~Alday and J.~M.~Maldacena,
  %``Gluon scattering amplitudes at strong coupling,''
  JHEP {\bf 0706} (2007) 064
  [arXiv:0705.0303 [hep-th]].
  %%CITATION = JHEPA,0706,064;%%

%\cite{Drummond:2007au}
\bibitem{Drummond:2007au}
  J.~M.~Drummond, J.~Henn, G.~P.~Korchemsky and E.~Sokatchev,
  %``Conformal Ward identities for Wilson loops and a test of the duality with
  %gluon amplitudes,''
  Nucl.\ Phys.\  B {\bf 826} (2010) 337
  [arXiv:0712.1223 [hep-th]].
  %%CITATION = NUPHA,B826,337;%%

%%%%%% REVIEWS

%\cite{Alday:2008yw}
\bibitem{Alday:2008yw}
  L.~F.~Alday and R.~Roiban,
  %``Scattering Amplitudes, Wilson Loops and the String/Gauge Theory
  %Correspondence,''
  Phys.\ Rept.\  {\bf 468} (2008) 153
  [arXiv:0807.1889 [hep-th]].
  %%CITATION = PRPLC,468,153;%%

%\cite{Henn:2009bd}
\bibitem{Henn:2009bd}
  J.~M.~Henn,
  %``Duality between Wilson loops and gluon amplitudes,''
  Fortsch.\ Phys.\  {\bf 57} (2009) 729
  [arXiv:0903.0522 [hep-th]].
  %%CITATION = FPYKA,57,729;%%

%%%%%%%%%%%


%%%%%%%%%%%
% Higgs regulator discussed in string theory:


%\cite{Kawai:2007eg}
\bibitem{Kawai:2007eg}
  H.~Kawai and T.~Suyama,
  %``Some Implications of Perturbative Approach to AdS/CFT Correspondence,''
  Nucl.\ Phys.\  B {\bf 794} (2008) 1
  [arXiv:0708.2463 [hep-th]].
  %%CITATION = NUPHA,B794,1;%%

%\cite{McGreevy:2008zy}
\bibitem{McGreevy:2008zy}
  J.~McGreevy and A.~Sever,
  %``Planar scattering amplitudes from Wilson loops,''
  JHEP {\bf 0808} (2008) 078
  [arXiv:0806.0668 [hep-th]].
  %%CITATION = JHEPA,0808,078;%%

%\cite{Berkovits:2008ic}
\bibitem{Berkovits:2008ic}
  N.~Berkovits and J.~Maldacena,
  %``Fermionic T-Duality, Dual Superconformal Symmetry, and the Amplitude/Wilson
  %Loop Connection,''
  JHEP {\bf 0809} (2008) 062
  [arXiv:0807.3196 [hep-th]].
  %%CITATION = JHEPA,0809,062;%%

%%%%%%%%%%%
% Higgs regulator in field theory

%\cite{Schabinger:2008ah}
\bibitem{Schabinger:2008ah}
  R.~M.~Schabinger,
  %``Scattering on the Moduli Space of N=4 Super Yang-Mills,''
  arXiv:0801.1542 [hep-th].
  %%CITATION = ARXIV:0801.1542;%%

%\cite{Alday:2009zm}
\bibitem{Alday:2009zm}
  L.~F.~Alday, J.~M.~Henn, J.~Plefka and T.~Schuster,
  %``Scattering into the fifth dimension of N=4 super Yang-Mills,''
  JHEP {\bf 1001} (2010) 077
  [arXiv:0908.0684 [hep-th]].
  %%CITATION = JHEPA,1001,077;%%




%\cite{Bern:2006ew}
\bibitem{Bern:2006ew}
  Z.~Bern, M.~Czakon, L.~J.~Dixon, D.~A.~Kosower and V.~A.~Smirnov,
  %``The Four-Loop Planar Amplitude and Cusp Anomalous Dimension in Maximally
  %Supersymmetric Yang-Mills Theory,''
  Phys.\ Rev.\  D {\bf 75} (2007) 085010
  [arXiv:hep-th/0610248].
  %%CITATION = PHRVA,D75,085010;%%
  
%\cite{Cachazo:2006az}
\bibitem{Cachazo:2006az}
  F.~Cachazo, M.~Spradlin and A.~Volovich,
  %``Four-Loop Cusp Anomalous Dimension From Obstructions,''
  Phys.\ Rev.\  D {\bf 75} (2007) 105011
  [arXiv:hep-th/0612309].
  %%CITATION = PHRVA,D75,105011;%%


  
  
%\cite{Henn:2010bk}
\bibitem{Henn:2010bk}
  J.~M.~Henn, S.~G.~Naculich, H.~J.~Schnitzer and M.~Spradlin,
  %``Higgs-regularized three-loop four-gluon amplitude in N=4 SYM:
  %exponentiation and Regge limits,''
  JHEP {\bf 1004} (2010) 038
  [arXiv:1001.1358 [hep-th]].
  %%CITATION = JHEPA,1004,038;%%


%\cite{Henn:2010ir}
\bibitem{Henn:2010ir}
  J.~M.~Henn, S.~G.~Naculich, H.~J.~Schnitzer and M.~Spradlin,
  %``More loops and legs in Higgs-regulated N=4 SYM amplitudes,''
  arXiv:1004.5381 [hep-th].
  %%CITATION = ARXIV:1004.5381;%%
  
%%%%%%%%%%%


  
%\cite{Gorsky:2009dr}
\bibitem{Gorsky:2009dr}
  A.~Gorsky and A.~Zhiboedov,
  %``Aspects of the N=4 SYM amplitude -- Wilson polygon duality,''
  arXiv:0911.3626 [hep-th].
  %%CITATION = ARXIV:0911.3626;%%

%\cite{Mason:2010pg}
\bibitem{Mason:2010pg}
  L.~Mason and D.~Skinner,
  %``Amplitudes at Weak Coupling as Polytopes in AdS_5,''
  arXiv:1004.3498 [hep-th].
  %%CITATION = ARXIV:1004.3498;%%

%\cite{Davydychev:1997wa}
\bibitem{Davydychev:1997wa}
  A.~I.~Davydychev and R.~Delbourgo,
  %``A geometrical angle on Feynman integrals,''
  J.\ Math.\ Phys.\  {\bf 39} (1998) 4299
  [arXiv:hep-th/9709216].
  %%CITATION = JMAPA,39,4299;%%

%\cite{Hodges:2010kq}
\bibitem{Hodges:2010kq}
  A.~Hodges,
  %``The box integrals in momentum-twistor geometry,''
  arXiv:1004.3323 [hep-th].
  %%CITATION = ARXIV:1004.3323;%%

%\cite{ArkaniHamed:2009dn}
\bibitem{ArkaniHamed:2009dn}
  N.~Arkani-Hamed, F.~Cachazo, C.~Cheung and J.~Kaplan,
  %``A Duality For The S Matrix,''
  JHEP {\bf 1003} (2010) 020
  [arXiv:0907.5418 [hep-th]].
  %%CITATION = JHEPA,1003,020;%%




%%%%%%%%%%%%%%%%
% cusp anomalous dimension (of a light-like cusp)

%\cite{Korchemskaya:1992je}
\bibitem{Korchemskaya:1992je}
  I.~A.~Korchemskaya and G.~P.~Korchemsky,
  %``On lightlike Wilson loops,''
  Phys.\ Lett.\  B {\bf 287} (1992) 169.
  %%CITATION = PHLTA,B287,169;%%





%%%%%%%%%%%%%%%%
% IR divergences in massive amplitudes/form factors:


%\cite{Korchemsky:1988hd}
\bibitem{Korchemsky:1988hd}
  G.~P.~Korchemsky,
  %``Sudakov Form-Factor In QCD,''
  Phys.\ Lett.\  B {\bf 220} (1989) 629.
  %%CITATION = PHLTA,B220,629;%%


%\cite{Mitov:2006xs}
\bibitem{Mitov:2006xs}
  A.~Mitov and S.~Moch,
  %``The singular behavior of massive QCD amplitudes,''
  JHEP {\bf 0705} (2007) 001
  [arXiv:hep-ph/0612149].
  %%CITATION = JHEPA,0705,001;%%




%\cite{Drummond:2006rz}
\bibitem{Drummond:2006rz}
  J.~M.~Drummond, J.~Henn, V.~A.~Smirnov and E.~Sokatchev,
  %``Magic identities for conformal four-point integrals,''
  JHEP {\bf 0701} (2007) 064
  [arXiv:hep-th/0607160].
  %%CITATION = JHEPA,0701,064;%%


%\cite{Drummond:2008vq}
\bibitem{Drummond:2008vq}
  J.~M.~Drummond, J.~Henn, G.~P.~Korchemsky and E.~Sokatchev,
  %``Dual superconformal symmetry of scattering amplitudes in N=4
  %super-Yang-Mills theory,''
  Nucl.\ Phys.\  B {\bf 828} (2010) 317
  [arXiv:0807.1095 [hep-th]].
  %%CITATION = NUPHA,B828,317;%%


%\cite{Jevicki:1998qs}
\bibitem{Jevicki:1998qs}
  A.~Jevicki, Y.~Kazama and T.~Yoneya,
  %``Quantum metamorphosis of conformal transformation in D3-brane  Yang-Mills
  %theory,''
  Phys.\ Rev.\ Lett.\  {\bf 81} (1998) 5072
  [arXiv:hep-th/9808039].
  %%CITATION = PRLTA,81,5072;%%


%%%%%%%%%%%%%%%
% pseudoconformal papers:

%\cite{Drummond:2007aua}
\bibitem{Drummond:2007aua}
  J.~M.~Drummond, G.~P.~Korchemsky and E.~Sokatchev,
  %``Conformal properties of four-gluon planar amplitudes and Wilson loops,''
  Nucl.\ Phys.\  B {\bf 795} (2008) 385
  [arXiv:0707.0243 [hep-th]].
  %%CITATION = NUPHA,B795,385;%%
  
  %\cite{Nguyen:2007ya}
\bibitem{Nguyen:2007ya}
  D.~Nguyen, M.~Spradlin and A.~Volovich,
  %``New Dual Conformally Invariant Off-Shell Integrals,''
  Phys.\ Rev.\  D {\bf 77} (2008) 025018
  [arXiv:0709.4665 [hep-th]].
  %%CITATION = PHRVA,D77,025018;%%

%\cite{Bern:2007ct}
\bibitem{Bern:2007ct}
  Z.~Bern, J.~J.~M.~Carrasco, H.~Johansson and D.~A.~Kosower,
  %``Maximally supersymmetric planar Yang-Mills amplitudes at five loops,''
  Phys.\ Rev.\  D {\bf 76} (2007) 125020
  [arXiv:0705.1864 [hep-th]].
  %%CITATION = PHRVA,D76,125020;%%

  
%\cite{Bern:2008ap}
\bibitem{Bern:2008ap}
  Z.~Bern, L.~J.~Dixon, D.~A.~Kosower, R.~Roiban, M.~Spradlin, C.~Vergu and A.~Volovich,
  %``The Two-Loop Six-Gluon MHV Amplitude in Maximally Supersymmetric Yang-Mills
  %Theory,''
  Phys.\ Rev.\  D {\bf 78} (2008) 045007
  [arXiv:0803.1465 [hep-th]].
  %%CITATION = PHRVA,D78,045007;%%


%\cite{Boels:2010mj}
\bibitem{Boels:2010mj}
  R.~H.~Boels,
  %``No triangles on the moduli space of maximally supersymmetric gauge
  %theory,''
  arXiv:1003.2989 [hep-th].
  %%CITATION = ARXIV:1003.2989;%%


%%%%%%%%%%%%%%%
% MB stuff


%\cite{Czakon:2005rk}
\bibitem{Czakon:2005rk}
  M.~Czakon,
  %``Automatized analytic continuation of Mellin-Barnes integrals,''
  Comput.\ Phys.\ Commun.\  {\bf 175} (2006) 559
  [arXiv:hep-ph/0511200].
  %%CITATION = CPHCB,175,559;%%

\bibitem{MBasymptotics}
  M.~Czakon, \\
  http://projects.hepforge.org/mbtools/.


\end{thebibliography}
\end{document}